\documentclass[aps,prl,preprint]{revtex4}
\usepackage{graphicx}
\usepackage{bm}
\usepackage{subfigure}
\usepackage{graphicx}
\begin{document}
\title{A Quantum Fluid Description of the Free Electron Laser}
\author{R. Bonifacio$^1$, N. Piovella$^2$, G.R.M Robb$^3$ \& A. Serbeto$^4$}
\affiliation{$^1$I.N.F.N. Sezione di Milano, Via
Celoria 16, Milano I-20133,Italy \\
$^2$Dipartimento di Fisica, Universita degli Studi di Milano, via Celoria 16, 20133 Milano\\
$^3$Scottish Universities Physics Alliance (SUPA), Department of Physics, University of Strathclyde, \\
107 Rottenrow, Glasgow, G4 0NG, United Kingdom \\
$^4$ Instituto de F\'{i}sica, Universidade Federal Fluminense, 24210-340 Rio de Janeiro, Brazil
}
\date{\today}
\begin{abstract}
Using the Madelung transformation we show that in a quantum Free
Electron Laser (QFEL) the beam obeys the equations of a quantum
fluid in which the potential is the classical potential plus a
quantum potential. The classical limit is shown explicitly.
\end{abstract}

\maketitle

\section{Introduction}
\label{intro}

In the quantum FEL model, the electron beam is described as a
macroscopic matter-wave\cite{EPL,NIMA,PRST,Plasmas}. When slippage due to
the difference between the light and electron velocities is
neglected, the electron beam-wave interaction is described by the
following equations for the dimensionless radiation amplitude
$A(\bar z)$ and the matter wave field $\Psi(\theta,\bar z)$
\cite{PREP}:
\begin{eqnarray}
i\frac{\partial \Psi (\theta ,\bar{z})}{\partial \bar{z}}
&=&-\frac{1}{ 2\bar{\rho}}\frac{\partial ^{2}}{\partial \theta
^{2}}\Psi (\theta, \bar{z})-i\bar{\rho}\left[
A(\bar{z})e^{i\theta }-\mathrm{
c.c.}\right] \Psi (\theta ,\bar{z}) \label{MS1}\\
\frac{d A(\bar{z})}{d \bar{z}} &=&\int_{0}^{2\pi }\;d\theta |\Psi
(\theta ,\bar{z})|^{2}e^{-i\theta }+i \delta
A(\bar{z})\label{MS2}.
\end{eqnarray}
The electron beam is therefore described by a Schr\"{o}dinger
equation for a matter-wave field $\Psi$ in a self-consistent
pendulum potential proportional to $A$, where
$|A|^2=|a|^2/(N\bar\rho)$, $|a|^2$ is the average number of
photons in the interaction volume $V$, and $|\Psi|^2$ is the
space-time dependent electron density, normalized to unity. In
Eqs. (\ref{MS1}) and (\ref{MS2}) we have adopted the universal
scaling used in the classical FEL theory \cite{BPN,SASE,NC},
\textit{i.e.} $\theta=(k+k_w)z-ckt$ is the electron phase, where
$k_w=2\pi/\lambda_w$ and $k=\omega/c=2\pi/\lambda$ are the wiggler
and radiation wavenumbers, $\bar z=z/L_g$ is the dimensionless
wiggler length, $L_g=\lambda_w/4\pi\rho$ is the gain length,
$\rho=\gamma_r^{-1}(a_w/4ck_w)^{2/3}(e^2n/m\epsilon_0)^{1/3}$ is
the classical FEL parameter,
$\gamma_r=\sqrt{(\lambda/2\lambda_w)(1+a_w^2)}$ is the resonant
energy in $mc^2$ units, $a_w$ is the wiggler parameter and $n$ is
the electron density. Finally, $\bar
p=(\gamma-\gamma_0)/\rho\gamma_0$ is the dimensionless electron
momentum and $\delta=(\gamma_0-\gamma_r)/\rho\gamma_0$ is the
detuning parameter, where $\gamma_0\approx\gamma_r$ is the initial
electron energy in $mc^2$ units.

Whereas the classical FEL equations in the above universal scaling
do not contain any explicit parameter (see ref. \cite{NC}), the
quantum FEL equations (\ref{MS1}) and (\ref{MS2}) depend on the
quantum FEL parameter
\begin{equation}\label{rhobar}
    \bar\rho=\left(\frac{mc\gamma_r}{\hbar k}\right)\rho.
\end{equation}
From the definition of $A$, it follows that
$\bar\rho|A|^2=|a|^2/N$ is the average number of photons emitted
per electron. Hence, since in the classical steady-state high-gain
FEL $A$ reaches a maximum value of the order of unity, $\bar\rho$
represents the maximum number of photons emitted per electron, and
the classical regime occurs for $\bar\rho\gg 1$. Note also that in
Eq. (\ref{MS1}) $\bar\rho$ appears as a ``mass'' term, so one
expects a classical limit when the mass is large. As we shall see,
when $\bar\rho < 1$ the dynamical behavior of the system changes
substantially from a classical to a quantum regime.

\section{Quantum Fluid Description}
\label{fluid}
We now perform a Madelung-like transformation \cite{Messiah}, writing the
wavefunction as
\[
\Psi = R \exp{(i \bar{\rho} S)}
\]
which allows us to rewrite the Maxwell-Schrodinger equations,
eq.~(\ref{MS1}) and (\ref{MS2}), as a system of quantum fluid
equations
\begin{eqnarray}
\frac{\partial R}{\partial \bar{z}} &=& -\frac{\partial
R}{\partial \theta} \frac{\partial S}{\partial \theta} -
\frac{R}{2}\frac{\partial^2 S}{\partial \theta^2}
\label{Mad1} \\
\frac{\partial S}{\partial \bar{z}} &=& -\frac{1}{2}
\left(\frac{\partial S} {\partial \theta} \right)^2
- V \left( \theta, \bar{z}\right)\label{Mad2} \\
\frac{d A}{d \bar{z}} &=& \int_0^{2 \pi} R^2 e^{-i \theta} \; d
\theta + i \delta A\label{Mad3}
\end{eqnarray}
where the potential, $V$ in eq.~(\ref{Mad2}) is defined as the sum of a classical
term and a quantum term i.e.
\[
V \left( \theta, \bar{z} \right) = V_C + V_Q
\]
where
\begin{equation}
V_C = -i \left ( A e^{i \theta} - c.c. \right) \label{pot_c}
\end{equation}
is the classical component of the potential and
\begin{equation}
V_Q = -\frac{1}{2 \bar{\rho}^2 R} \frac{\partial^2 R}{\partial
\theta^2} \label{pot_q}
\end{equation}
is the quantum component of the potential, which becomes negligible as $\bar{\rho}
\rightarrow \infty$.

Defining fluid density and velocity variables
\[
n = R^2=|\Psi|^2\;\;,\;\;\;\;u =  \frac{\partial S}{\partial
\theta}
\]
we can also rewrite Eq.~(\ref{Mad1})-(\ref{Mad3}) in an
alternative fluid form as
\begin{eqnarray}
\frac{\partial n}{\partial \bar{z}} + \frac{\partial}{\partial \theta}
\left( n u \right)&=& 0 \label{Mad_cont} \\
\frac{\partial u}{\partial \bar{z}} + u \frac{\partial u}{\partial \theta} &=&
- \frac{\partial V}{\partial \theta} \label{Mad_force} \\
\frac{d A}{d \bar{z}} &=& \int_0^{2 \pi} n e^{-i \theta} \; d
\theta + i \delta A \label{Mad_fld}.
\end{eqnarray}
It can be seen that Eq.~(\ref{Mad_cont}) is a continuity equation
and Eq.~(\ref{Mad_force}) is a Newton-like equation for a fluid.
Note that integrating Eq.~(\ref{Mad_cont}) with respect to
$\theta$, then the normalization condition becomes
\[
\int_0^{2 \pi} n(\theta,\bar z) d \,\theta = 1,
\]
which is satisfied if $n$ and $u$ are periodic functions of
$\theta$ between $0$ and $2 \pi$.

A straightforward calculation shows that
Eq.(\ref{Mad_cont})-(\ref{Mad_fld}) admit two constants of motion,
\begin{equation}
\langle\bar p\rangle+|A|^2={\cal C}_1 \label{C1}
\end{equation}
and
\begin{equation}\label{H}
\frac{\langle\bar p^2\rangle}{2}-i(Ab^*-c.c.)-\delta|A|^2={\cal
C}_2
\end{equation}
where $\langle\bar p\rangle=\langle u\rangle=\int_0^{2\pi}d\theta
nu$ is the average momentum, $\langle\bar p^2\rangle=\langle
u^2+2V_Q\rangle=\int_0^{2\pi}d\theta n(u^2+2V_Q)$ is the momentum
variance and
\[
b=\int_0^{2 \pi} n e^{-i \theta} \; d \theta
\]
is the bunching. These constants of motion are well-known in the
classical FEL model \cite{NC} and describe energy conservation and
a gain-spread relation. Notice the quantum contribution to the
momentum variance proportional to the average quantum potential.

%\section{Steady-State Solution of Fluid Equations}
%\textsl{[I do not understand this section,, what does it means and
%what is its importance? It exists such steady-state solution?
%Nicola \\
%I also don't think this is so important. It doesn't really fit with the
% rest of the paper and could be removed without affecting it - Gordon]}

%We now consider a steady-state solution of the quantum fluid
%equations (eq.~(\ref{Mad1}..\ref{Mad3})). Following \cite{Messiah}
%we assume a stationary state with energy $E$ such that
%\[
%\frac{\partial S}{\partial \bar{z}} = - E \;\;\;,\;\;\;\frac{\partial R}
%{\partial \bar{z}} = 0 \;\;\;,\;\;\;\frac{d A}{d \bar{z}}=0.
%\]
%In this steady-state limit, eq.(\ref{Mad1}..\ref{Mad2}) become
%\begin{eqnarray}
%\frac{1}{2} \left( \frac{d S}{d \theta} \right)^2 = E - V(\theta) \label{Mad1_ss} \\
%\frac{\partial}{\partial \theta}
%\left( R^2 \frac{\partial S}{\partial \theta}\right) = 0 \label{Mad2_ss}
%\end{eqnarray}

\section{Fourier Expansion and Linear Analysis}

If $R$ and $S$ are periodic functions of $\theta$, they can be
expanded in a Fourier series:
\begin{eqnarray}
  R(\theta,\bar z) &=& \sum_m r_m(\bar z)
  e^{im\theta},\label{rn}\\
  S(\theta,\bar z) &=& \sum_n s_n(\bar z) e^{in\theta} \label{sn}
\end{eqnarray}
with $r_{-m}=s^{*}_m$ and $s_{-m}=r^{*}_m$, since $R$ and $S$ are
real variables. Multiplying Eq.(\ref{Mad2}) by $R$ and using
(\ref{rn}) and (\ref{sn}) in Eqs.(\ref{Mad1})-(\ref{Mad3}), we
obtain:
\begin{eqnarray}
  \sum_m r_{k-m}\frac{ds_{m}}{d\bar z} &=&
  -\frac{1}{2}\sum_{m,n}
  n(n-m)r_{k-m}s_ns^*_{n-m}+i(Ar_{k-1}-A^*r_{k+1})-\frac{k^2}{2\bar\rho^2}r_k
  \label{Mad1:n} \\
   \frac{dr_{k}}{d\bar z}&=&
   \frac{1}{2}\sum_m (k^2-m^2)r_m s_{m-k}^*
   \label{Mad2:n}\\
   \frac{dA}{d\bar z}&=&\sum_m r_mr^{*}_{m-1}+i\delta A
   \label{Mad3:n}
\end{eqnarray}
Eqs.(\ref{Mad1:n})- (\ref{Mad3:n}) are our working equations which
can be numerically solved as it will be shown elsewhere.

Eqs.(\ref{Mad1:n})-(\ref{Mad3:n}) admit an equilibrium solution
with no field ($A=0$) and unbunched electron beam ($n=1/2\pi$,
i.e. $r_n=\delta_{n0}$ and $s_n=0$). Linearizing
Eqs.(\ref{Mad1:n})-(\ref{Mad3:n}) around this equilibrium in the
first order of the variables $A$, $r_1$ and $s_1$, we obtain:
\begin{eqnarray}
   \frac{dA}{d\bar z}&=& 2r_1+i\delta A\\
   \frac{dr_{1}}{d\bar z}&=& \frac{s_1}{2}\\
  \frac{ds_{1}}{d\bar z}&=& iA -\frac{1}{2\bar\rho^2}r_1
\end{eqnarray}
Looking for solutions proportional to $\exp(i\lambda \bar z)$, we
obtain the well-known cubic equation of the quantum FEL \cite{EPL}
\begin{equation}\label{cubic}
    (\lambda-\delta)\left(\lambda^2-\frac{1}{4\bar\rho^2}\right)+1=0.
\end{equation}
which reduces to the classical dispersion relation in the limit
$\bar\rho>>1$.

\section{Conclusions}
It has been shown that the quantum FEL model can be rewritten in a
form where the electron beam is described a quantum fluid coupled
to the electromagnetic field. The evolution of the quantum fluid
is determined by a self-consistent potential which consists of a
classical and quantum contribution. In the limit where $\bar{\rho}
\gg 1$ the quantum contribution to the potential becomes
negligible and the force equation reduces to that of a Newtonian
fluid. Using a Fourier expansion, linear stability analysis of
these quantum fluid equations produced a dispersion relation
identical to that derived from the Schrodinger equation. These
results show that there are interesting connections between the
quantum FEL and quantum plasma instabilities.

\end{document}